\documentclass[reprint,amsmath,amssymb,aps,pra]{revtex4-1}
\usepackage{graphicx,color}
\usepackage{bm}
\usepackage{braket}
\usepackage{mathtools}
\usepackage{amsfonts}
\usepackage{ulem}
\usepackage[breaklinks,colorlinks=true,linkcolor=blue,urlcolor=blue,citecolor=magenta]{hyperref}

\begin{document}
\title{Systematic construction of topological flat-band models by molecular-orbital representation}
\author{Tomonari Mizoguchi}
\affiliation{Department of Physics, University of Tsukuba, Tsukuba, Ibaraki 305-8571, Japan}
\email{mizoguchi@rhodia.ph.tsukuba.ac.jp}
\author{Yasuhiro Hatsugai}
\affiliation{Department of Physics, University of Tsukuba, Tsukuba, Ibaraki 305-8571, Japan}

\date{\today}
\begin{abstract}
On the basis of the ``molecular-orbital" representation which describes generic flat-band models, 
we propose a systematic way to construct a class of flat-band models with finite-range hoppings that have topological natures. 
In these models, the topological natures are encoded not into the flat band itself 
but into the dispersive bands touching the flat band. 
Such a band structure may become a source of exotic phenomena arising from the combination of flat bands, topology and correlations.
\end{abstract}

\maketitle
\section{Introduction}
Interplay among flat bands, topology, and electron-electron correlation gives rise to intriguing physics.
A typical example is the fractional quantum Hall effect (FQHE)~\cite{Tsui1982}.
In the two-dimensional electron gas, 
the formation of completely flat Landau levels occurs due to an external magnetic field, 
and the electron-electron correlation leads to the emergence 
of fractionalized quasi-particles composed of the electrons being attached to the flux~\cite{Jain1989}.  

Novel aspects of FQHE have still been studied actively~\cite{Katsura2010,Green2010,Tang2011,Sun2011,Neupert2011,Sheng2011,Regnault2011,Qi2011,Liu2012,Takayoshi2013,Bergholtz2013,Lee2014,Udagawa2014,Bergholtz2015,Behrmann2016,CHLee2016,CHLee2017,Son2015,Kane2002,Teo2014,Fuji2019,Yoshida2019_2}. 
One of the prominent examples is the realization of the FQHE without external magnetic fields,
which is called fractional Chern insulators (FCIs)~\cite{Katsura2010,Green2010,Tang2011,Sun2011,Neupert2011,Sheng2011,Regnault2011,Qi2011,Liu2012,Takayoshi2013,Bergholtz2013,Lee2014,Udagawa2014,Bergholtz2015,Behrmann2016,CHLee2016,CHLee2017}.
A key ingredient to realize FCIs is the exact or nearly flat bands with finite Chern number. 
Together with the theoretical developments, 
candidate materials for such phenomena have been intensively explored. 
Examples include metal organic frameworks with ions having strong spin-orbit coupling~\cite{Liu2013,Yamada2016}, and twisted bilayer graphene~\cite{Cao2018,Cao2018_2,Koshino2018,Spanton2018,Hejazi2019,Hejazi2019_2}. 

To study exotic phases due to the combination of flat band, topology, and correlations, 
simple tight-binding models having flat bands
are expected to provide a good starting point~\cite{Guo2009,Weeks2010,Pal2018,Jana2019,Bhattacharya2019,Lima2019}. 
Such models are defined on a class of Lieb lattices~\cite{Lieb1989,Sutherland1986} 
and line graphs~\cite{Mielke1991} 
and their variants~\cite{Miyahara2005}.
However, implementation of the topologically nontrivial structures 
to those well-known flat-band models, 
such as adding spin-orbit couplings, 
often leads to finite dispersion of flat bands~\cite{Bergholtz2013,Bergholtz2015,Guo2009,Kurita2011,Bolens2018}. 
For this reason, most of the theoretical studies of FCIs have been carried out on 
the models which have nearly flat bands with nontrivial topological natures.
In those models, the flatness of the bands is controlled quantitatively~\cite{Udagawa2014,CHLee2016,CHLee2017}.  

In this paper, 
we introduce a different approach to construct ``topological flat-band models".
Our models have finite-range hoppings and exact flat bands.
In such models, it was proved on the basis of $K$-theory that flat bands must not have a finite Chern number~\cite{Chen2014}.  
However, it is possible to construct the models whose dispersive bands are topologically nontrivial
and they have touching points with the flat bands. 
Such models will serve as another platform for studying the interplay among flat bands, topology, and electron correlations. 

\begin{figure}[t]
\centering
\includegraphics[clip,width=0.9\linewidth]{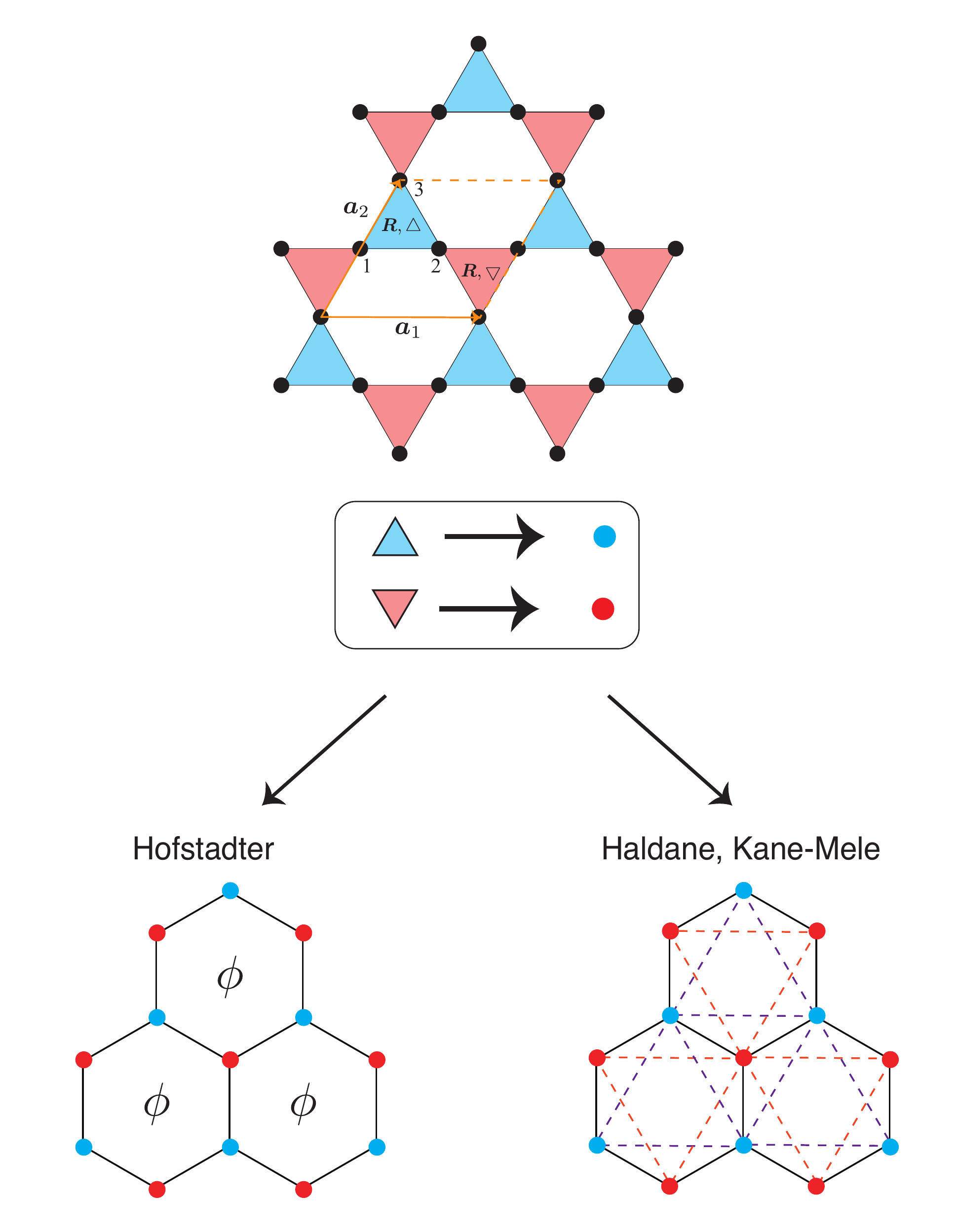}
\caption{Schematic figure for the construction of the tight-binding models considered in this paper. } 
\label{fig1}
\end{figure}

The key strategy to construct such models is to make use of the 
``molecular-orbital" (MO) representation, which was developed in the prior works~\cite{Hatsugai2011,Hatsugai2015,Mizoguchi2019}.
In this representation, we describe tight-binding models having flat bands
using non-orthogonal basis composed of a small number of atomic orbitals.
For the line graphs, which we will consider in this paper, 
the MOs are usually defined on the dual lattice, e.g., a honeycomb lattice for a kagome lattice.
Since there are a variety of examples of topological models 
defined on a honeycomb lattice,
one may simply implement such models for the molecular orbitals, and recast them into the original kagome lattice
(see Fig.~\ref{fig1} for the schematic of the construction of such models).
Then, the topologically nontrivial bands and the exact flat band coexist in the models thus obtained, as we will show.

The rest of this paper is organized as follows.
In Sec.~\ref{sec:MOrep}, we explain a method of a systematic construction of topological flat-band models
based on the MO representation.
Then, our main results are illustrated in Sec.~\ref{sec:result}, where we show three examples of 
topological flat-band models composed of MOs. 
In Sec.~\ref{sec:summary} , we present a summary of this paper.   
The Appendix~\ref{sec:pyrochlore} is devoted to the three-dimensional model constructed by the same method.

\section{Molecular-orbital representation of topological flat-band models \label{sec:MOrep}}
In this section, 
we explain how to construct topological flat-band models by using the MO representation. 
Throughout this paper, we focus on a tight-binding model on a kagome lattice of spinless or spinful fermions.
Application of the same strategy to other lattices is straightforward (see the Appendix~\ref{sec:pyrochlore} for an example on a pyrochlore lattice). 

On a kagome lattice, each site is labeled by the position of the unit cell 
$\bm{R} = r_1 \bm{a}_1 +r_2\bm{a}_2$,
and the sublattice $\eta = 1,2,3$; we use the abbreviated form $i = (\bm{R},\eta)$.
The annihilation and creation operators on $i$ are represented by $c_{i,\sigma}$, and $c_{i,\sigma}^{\dagger}$, respectively.
$\sigma = \uparrow, \downarrow$ labels the spin degrees of freedom for the spinful systems;
for the spinless systems, we simply omit this index. 

We first illustrate how the MO representation can yield the exact flat bands~\cite{Hatsugai2011,Mizoguchi2019}.
We consider the models written 
by the following non-orthogonal and 
unnormalized basis $C^{\bigtriangleup}_{\bm{R},\sigma}$ and $C^{\bigtriangledown}_{\bm{R},\sigma}$ 
which we call MOs:
\begin{eqnarray}
C^{\bigtriangleup}_{\bm{R},\sigma} = \gamma^{\bigtriangleup}_1 c_{\bm{R},1,\sigma}  
+ \gamma^{\bigtriangleup}_2 c_{\bm{R},2,\sigma} 
+ \gamma^{\bigtriangleup}_3 c_{\bm{R},3,\sigma}, \label{eq:kagome_mo_1}
\end{eqnarray}
and 
\begin{eqnarray}
C^{\bigtriangledown}_{\bm{R},\sigma} = \gamma^{\bigtriangledown}_1 c_{\bm{R} + \bm{a}_1,1,\sigma}  
+ \gamma^{\bigtriangledown}_2 c_{\bm{R},2,\sigma} 
+\gamma^{\bigtriangledown}_3 c_{\bm{R} +\bm{a}_1-\bm{a}_2 ,3 ,\sigma}, \label{eq:kagome_mo_2}
\end{eqnarray}
with $\gamma^{\bigtriangleup/\bigtriangledown}_{\eta} \in \mathbb{C}$.

These MOs are defined on the triangles, and thus they are placed on a honeycomb lattice. 
Now, let us consider the tight-binding models which can be written only by the MOs:
\begin{eqnarray}
\mathcal{H} =\sum_{\bm{R},\bm{R}^\prime} \bm{C}^{\dagger}_{\bm{R}} h_{\bm{R},\bm{R}^\prime} \bm{C}_{\bm{R}^\prime}, \label{eq:MOHam}
\end{eqnarray} 
where $\bm{C}_{\bm{R}} = \left( C^{\bigtriangleup}_{\bm{R},\uparrow}, C^{\bigtriangledown}_{\bm{R},\uparrow}, C^{\bigtriangleup}_{\bm{R},\downarrow}, C^{\bigtriangledown}_{\bm{R},\downarrow} \right)^{\mathrm{T}}$,
and $h_{\bm{R},\bm{R}^\prime}$ represents the Hamiltonian for the MOs which is defined on a honeycomb lattice. 
Using Eqs. (\ref{eq:kagome_mo_1}) and (\ref{eq:kagome_mo_2}), one can easily recast the model onto the original kagome lattice. 
The model thus obtained has an exact zero-energy flat band, since the projection from the original kagome sites onto the MOs causes the reduction of the degrees of freedom, and the kernel of the projection is enforced to have zero-energy~\cite{Hatsugai2011,Mizoguchi2019}.

The Hamiltonian of Eq. (\ref{eq:MOHam}) can be written in the momentum-space representation if $ h_{\bm{R},\bm{R}^\prime}$ 
has a translational symmetry, i.e., 
$h_{\bm{R},\bm{R}^\prime}$ depends only on 
$\bm{R}-\bm{R}^\prime$ and thus it can be written as 
$h_{\bm{R},\bm{R}^\prime} = h_{\bm{R}-\bm{R}^\prime}$.
In the original basis of a kagome lattice, it can be written as
\begin{eqnarray}
\mathcal{H} = & \sum_{\bm{k}} 
 \bm{c}^\dagger_{\bm{k} } \mathcal{H}_{\bm{k} } \bm{c}_{\bm{k} },
\end{eqnarray}
with 
$\bm{c}_{\bm{k}} = (c_{1,\bm{k},\uparrow}, c_{2,\bm{k},\uparrow},c_{3,\bm{k},\uparrow},c_{1,\bm{k},\downarrow},c_{2,\bm{k},\downarrow},c_{3,\bm{k},\downarrow})^{\mathrm{T}}$.
The Hamiltonian matrix 
$\mathcal{H}_{\bm{k} }$ has a form
\begin{eqnarray}
\mathcal{H}_{\bm{k} } = \Psi_{\bm{k}} h_{\bm{k}} \Psi^\dagger_{\bm{k}}, \label{eq:ham_gen}
\end{eqnarray}
where 
\begin{eqnarray}
h_{\bm{k}} = \sum_{\bm{R}} h_{\bm{R}} e^{-i \bm{k} \cdot \bm{R}},
\end{eqnarray}
is the Hamiltonian matrix for the MOs in the momentum-space representation,
and 
\begin{widetext}
\begin{eqnarray}
\Psi^\dagger_{\bm{k}} = \left(
\begin{array}{cccccc}
\gamma_1^\bigtriangleup & \gamma_2^\bigtriangleup & \gamma_3^\bigtriangleup & 0 & 0 & 0\\
\gamma_1^\bigtriangledown e^{i \bm{k} \cdot \bm{a}_1} & \gamma_2^\bigtriangledown & \gamma_3^\bigtriangledown e^{i \bm{k} \cdot (\bm{a}_1-\bm{a}_2)} & 0 & 0 & 0\\
 0 & 0 & 0 &\gamma_1^\bigtriangleup & \gamma_2^\bigtriangleup & \gamma_3^\bigtriangleup \\
 0 & 0 & 0 & \gamma_1^\bigtriangledown e^{i \bm{k} \cdot \bm{a}_1}  & \gamma_2^\bigtriangledown & \gamma_3^\bigtriangledown e^{i \bm{k} \cdot (\bm{a}_1-\bm{a}_2)}  \\
\end{array}
\right),\nonumber \\
\end{eqnarray}
\end{widetext}
is a matrix which maps the original basis on a kagome lattice to the MOs:
\begin{eqnarray}
\bm{C}_{\bm{k}}= \Psi_{\bm{k}}^\dagger   \bm{c}_{\bm{k}}.
\end{eqnarray}

The dispersion relation of the tight-binding Hamiltonian can be obtained by solving the eigenvalue equation,
\begin{eqnarray}
\mathcal{H}_{\bm{k}} \bm{u}_{n, \bm{k}} = \varepsilon_n (\bm{k})\bm{u}_{n, \bm{k}}, \label{eq:sch_gen}
\end{eqnarray}
where $\bm{u}_{n, \bm{k}}$ is the six-component column vector representing the $n$-th eigenvector
and $ \varepsilon_n$ is the $n$-th band energy. 
As $\mathcal{H}_{\bm{k}}$ has the form of Eq.~(\ref{eq:ham_gen}), 
two out of six eigenvalues are guaranteed to be zero \textit{regardless of the momentum $\bm{k}$}, 
meaning that they form flat bands.
This can be explained as follows:
As the matrix $\Psi^\dagger_{\bm{k}}$ is a $4 \times 6$ matrix, we have 
\begin{eqnarray}
\mathrm{dim} \left[ \mathrm{ker} \left( \Psi^\dagger_{\bm{k}}\right)\right] \geq 2. 
\end{eqnarray}
Let $\bm{v}_{1,\bm{k}}$ and $\bm{v}_{2,\bm{k}}$ be two linearly-independent vectors 
in the kernel of $\Psi^\dagger_{\bm{k}}$, i.e., $\Psi^\dagger_{\bm{k}} \bm{v}_{1/2, \bm{k}}  = 0$. 
Then, substituting these into the eigenvalue equation of Eq.~(\ref{eq:sch_gen}),
we have 
\begin{eqnarray}
\mathcal{H}_{\bm{k}} \bm{v}_{1/2, \bm{k}} 
= \left( \Psi_{\bm{k}} h_{\bm{k}} \Psi^\dagger_{\bm{k}}\right) 
\bm{v}_{1/2, \bm{k}} 
= 0.
\end{eqnarray}
This means that $\bm{v}_{1,\bm{k}}$ and $\bm{v}_{2,\bm{k}}$ are the zero-energy eigenvectors 
of $\mathcal{H}_{\bm{k}}$.

Henceforth, for simplicity, 
we set $(\gamma^{\bigtriangleup}_{1},\gamma^{\bigtriangleup}_{2},\gamma^{\bigtriangleup}_{3},\gamma^{\bigtriangledown}_{1},\gamma^{\bigtriangledown}_{2},\gamma^{\bigtriangledown}_{3} ) =(1,1,1,1,1,1)$.
For this choice, the quadratic band touching between the flat band and the dispersive band at the $\Gamma$ point
is enforced to occur, due to the reduction of the linear space spanned by the MOs~\cite{Bergman2008,Mizoguchi2019}.
Namely, at the $\Gamma$ point, $\Psi^\dagger_{\bm{k}=0}$ is given as  
\begin{eqnarray}
\Psi^\dagger_{\bm{k}=0} = \left(
\begin{array}{cccccc}
1 & 1 &1  & 0 & 0 & 0\\
1 & 1 &1 & 0 & 0 & 0\\
 0 & 0 & 0 &1& 1 & 1 \\
 0 & 0 & 0 &1& 1 & 1  \\
\end{array}
\right),\nonumber \\
\end{eqnarray}
whose kernel is four dimensional.   
This means that there has to be four zero modes; two of them come from the flat bands and the others from the dispersive bands.
If we choose $\gamma^{\bigtriangleup/\bigtriangledown}_{\eta}$ 
such that the vectors $(\gamma^{\bigtriangleup}_{1},\gamma^{\bigtriangleup}_{2},\gamma^{\bigtriangleup}_{3})$ and $(\gamma^{\bigtriangledown}_{1},\gamma^{\bigtriangledown}_{2},\gamma^{\bigtriangledown}_{3})$ are linearly independent with each other,
the band touching can be erased~\cite{Bilitewski2018,Mizoguchi2019}.

As we have seen, the emergence of the exact flat band and the band touching to the dispersive band at the $\Gamma$ point originate 
from $\Psi_{\bm{k}}^\dagger$, hence
$h_{\bm{R},\bm{R}^\prime}$ can be a generic tight-binding Hamiltonian.  
In the previous works, rather simple forms of $h_{\bm{R},\bm{R}^\prime}$ are considered. 
For instance, the nearest-neighbor hopping model on kagome and a breathing kagome lattices can be described by setting 
$h_{\bm{R},\bm{R}^\prime}$ as an ``on-site potential"~\cite{Hatsugai2011,Mizoguchi2019}.
Our strategy is to set $h_{\bm{R},\bm{R}^\prime}$ as well-known topological models defined on a honeycomb lattice, as we show in the next section. 
The models thus obtained have more or less complicated patterns of the hoppings on the original kagome lattice. 
Nevertheless, the hoppings are finite ranged, and the topological natures of $h_{\bm{R},\bm{R}^\prime}$ are indeed inherited by the dispersive bands.

\section{Examples of topological flat-band models on a kagome lattice \label{sec:result}}
\subsection{Molecular-orbital Hofstadter model}
\begin{figure}[t]
\centering
\includegraphics[bb=0 0 343 230,width=0.95\linewidth]{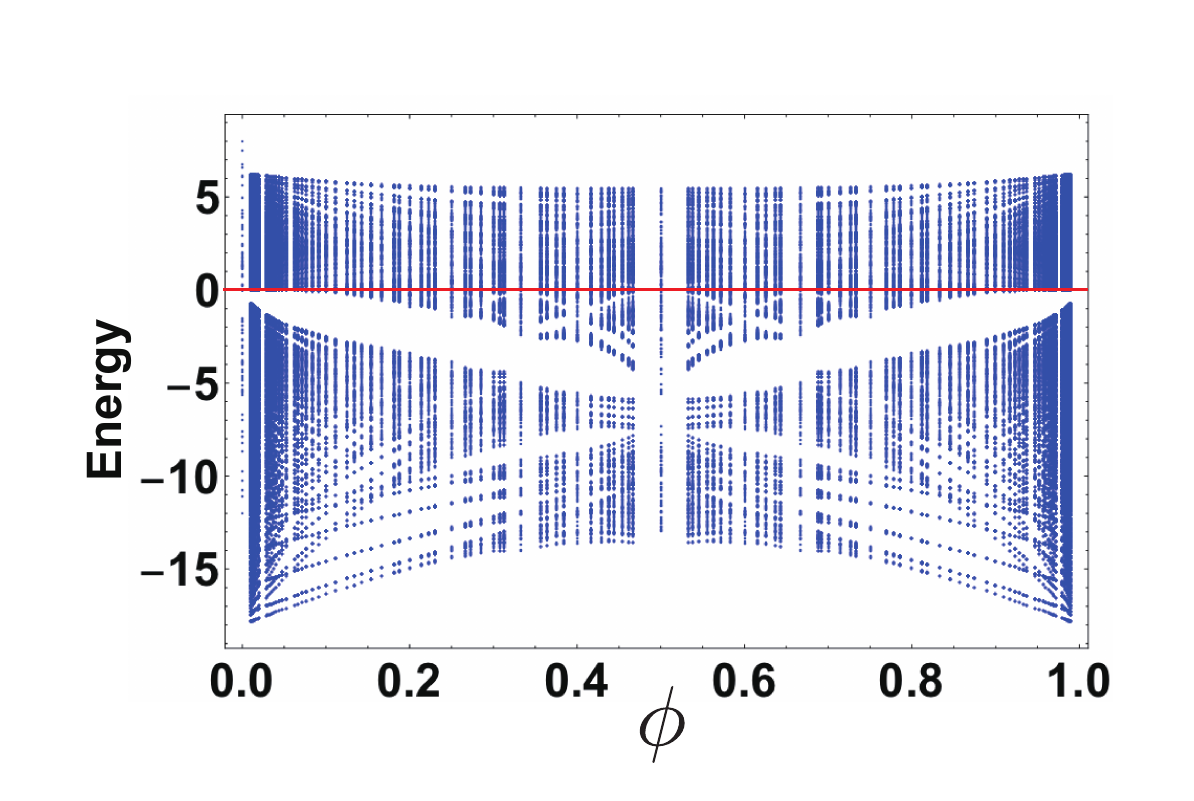}
\caption{The energy spectrum for the Hamiltonian of Eq. (\ref{eq:ham_hof}). 
A red line represents the degenerate zero-energy states, i.e., flat bands.} 
\label{fig2}
\end{figure}
\begin{figure}[t]
\centering
\includegraphics[clip,width=0.98\linewidth]{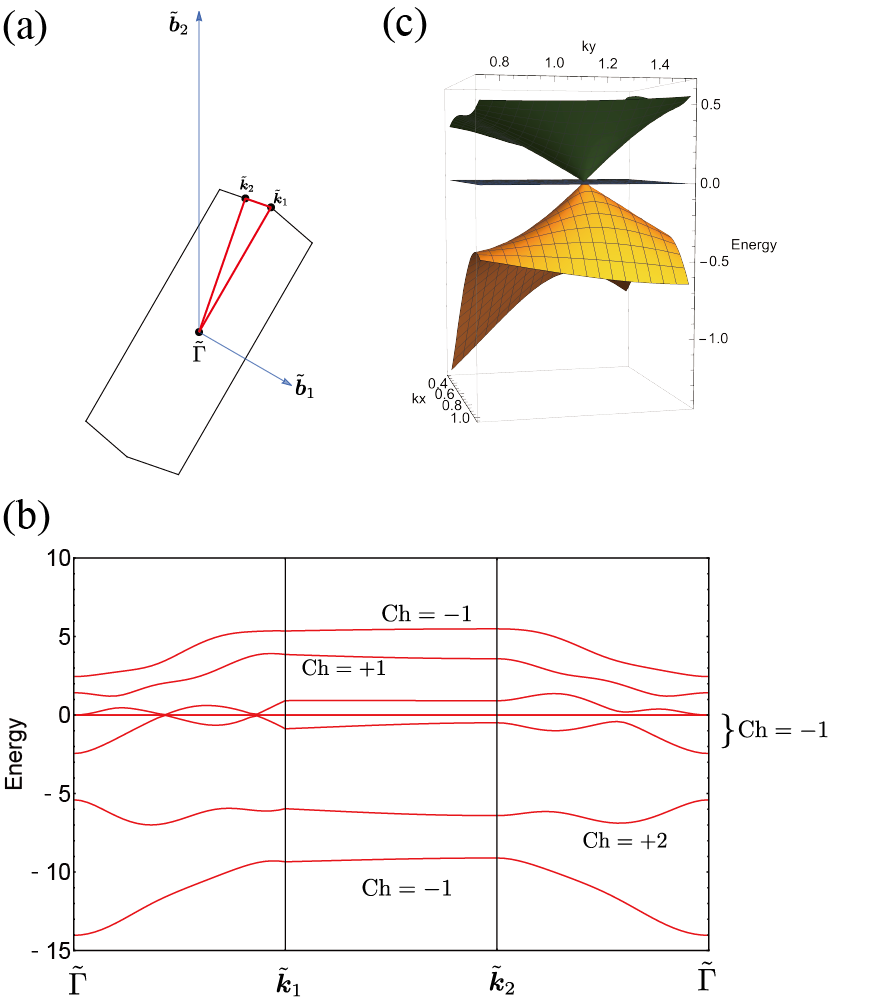}
\caption{
(a) The magnetic Brillouin zone for $\phi = 1/3$.
We set the magnetic lattice vectors as $\tilde{\bm{a}}_1 = 3\bm{a}_1$ and $\tilde{\bm{a}}_2 = \bm{a}_2$.
The corresponding reciprocal lattice vectors, $\tilde{\bm{b}}_1$ and $\tilde{\bm{b}}_2$, are denoted by blue arrows. 
The coordinates of the corners of the first Brillouin zones are 
$\tilde{\bm{k}}_1 =  \frac{7}{27} \left(3\tilde{\bm{b}}_1+2\tilde{\bm{b}}_2 \right)$ and $\tilde{\bm{k}}_2 =  \frac{1}{2} \left(\tilde{\bm{b}}_1+\tilde{\bm{b}}_2 \right)$.
(b) The band structure for $\phi=1/3$. 
The Chern numbers of the bands are indicated in the figure.
(c) A Dirac point formed by the third and the seventh band. 
At the Dirac point, they are also degenerated with three flat bands. 
} 
\label{fig2_2}
\end{figure}
The first model is the Hofstadter model~\cite{Hofstadter1976} for the molecular orbitals.
Here we consider the spinless fermions. 
The Hamiltonian reads
\begin{eqnarray}
\mathcal{H}(\phi) &=& 
\sum_{\bm{R}} -t C^{\bigtriangleup, \dagger}_{\bm{R}} 
\left( C^{\bigtriangledown}_{\bm{R}}  +e^{i 2\pi \phi r_1} C^{\bigtriangledown}_{\bm{R} - \bm{a}_1}  + C^{\bigtriangledown}_{\bm{R}-\bm{a}_1+\bm{a}_2} \right) \nonumber \\
&+& (\mathrm{H.c.}), \nonumber \\ \label{eq:ham_hof}
\end{eqnarray}
where 
$\phi = p/q$ with $p$ and $q$ are relatively prime numbers.
It should be emphasized that the model is different from the conventional 
Hofstadter model on a kagome lattice~\cite{Kimura2002} and its variants~\cite{Ohgushi2000,Maiti2019}.

In Fig.~\ref{fig2}, we show the energy spectrum as a function of $\phi$. 
The diagram resembles neither 
the honeycomb Hofstadter model~\cite{Rammal1985,Hatsugai2006} 
nor the kagome Hofstadter model~\cite{Kimura2002}.
Remarkably, the zero-energy modes with macroscopic degeneracy remain for any $\phi$. 
Note that the same behavior is also seen in the Hofstadter model on a Lieb lattice~\cite{Aoki1996} and a dice lattice~\cite{Vidal1998}.
\begin{figure}[t]
\centering
\includegraphics[clip,width=0.95\linewidth]{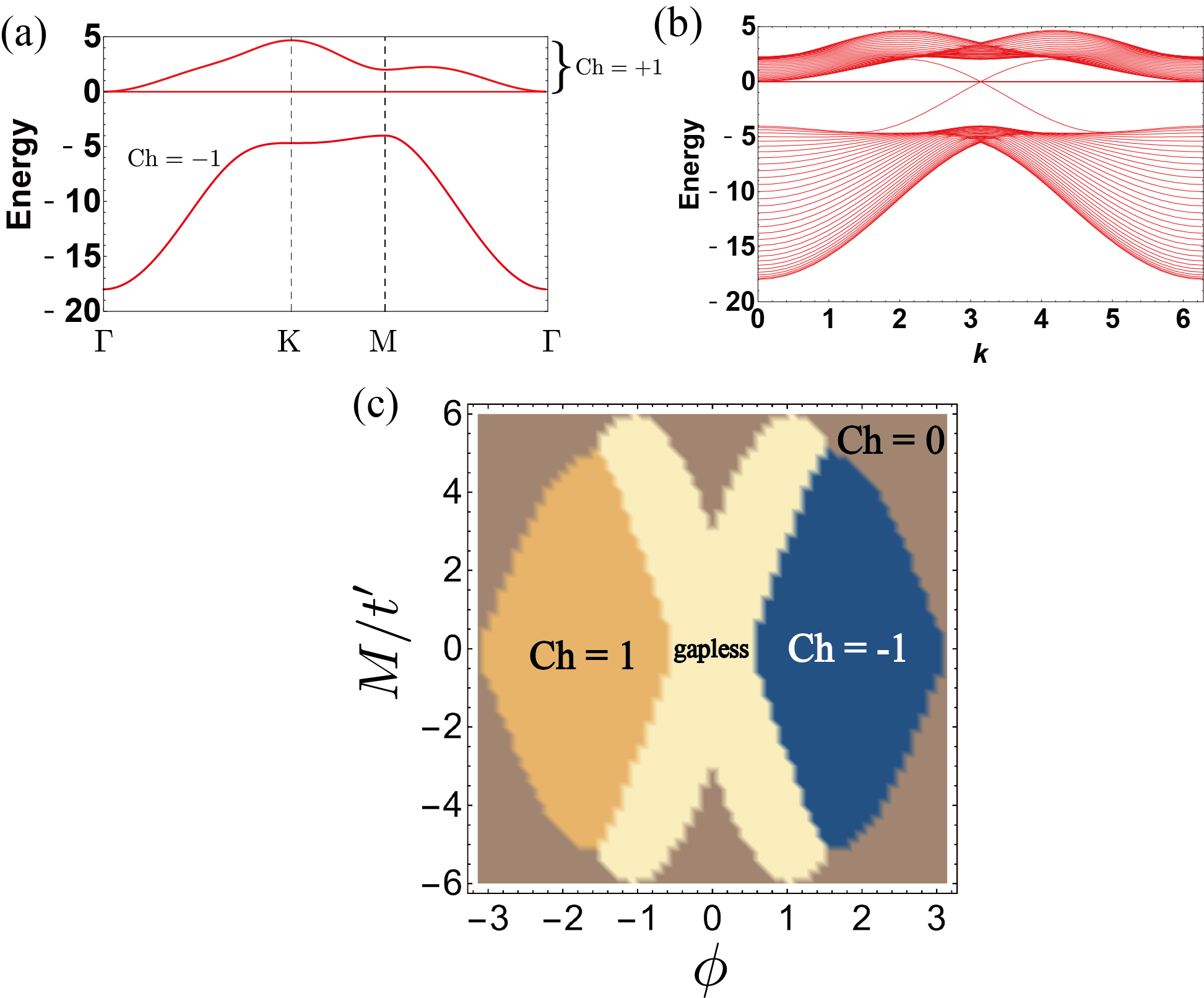}
\caption{ 
(a) The band structure of the MO Haldane model for $(t,t^\prime,M,\phi,\gamma) = (-1, -0.3, 0,\frac{\pi}{2},1)$.
The high-symmetry points in the Brillouin zone are $\Gamma = (0,0)$, $K=\left( \frac{4\pi}{3},0\right)$,
and $M=\left( \pi, \frac{\pi}{\sqrt{3}}\right)$.
(b) The band structure for the same parameter on a cylinder geometry. 
(c) A $M$-$\phi$ phase diagram of the MO Haldane model at $t^\prime = -0.3$. 
The orange, brown, and blue regions correspond to  
the Chern numbers of the lowest band is $1$, $0$, and $-1$, respectively. 
In the yellow region, the lowest dispersive band overlaps with the flat band.
} 
\label{fig4}
\end{figure}

To study the topological properties, let us look at the band structure for the specific choice of $p$ and $q$;
here we choose $p=1$, and $q=3$.
The band structure in the magnetic Brillouin zone [Fig.~\ref{fig2_2}(a)] and the Chern numbers 
computed numerically by using the method of Ref.~\onlinecite{Fukui2005}, 
are shown in Fig.~\ref{fig2_2}(b). 
We see that the flat bands, which have three-fold degeneracy,  
touch the dispersive band at the $\overline{\Gamma}$ point.
Furthermore,
we also find Dirac cones along the $\tilde{\Gamma}$-$\tilde{\bm{k}}_1$ line,
which are formed by the dispersive bands whose Dirac points degenerate 
with the flat band as well [Fig.~\ref{fig2_2}(c)].
These Dirac cones originate from those of the Hofstadter model on a honeycomb lattice.
Namely, the band touching between flat bands and dispersive bands occurs 
when $\mathrm{det} \hspace{.5mm} h_{\bm{k}}= 0$ or when $\mathrm{det} \hspace{.5mm}\Psi^\dagger_{\bm{k}}\Psi_{\bm{k}} = 0$~\cite{Mizoguchi2019}.
In the present model, $h_{\bm{k}}$ is nothing but a Hofstadter model on a honeycomb lattice, 
and it hosts Dirac cones where det $h_{\bm{k}} = 0$ is satisfied for certain values of $\phi$. 
Then, the multiplet of middle bands,
composed of two dispersive bands and three-fold-degenerate flat bands, 
has the Chern number $-1$ in total. 
This result indicates that it is possible to make the flat band touch the topologically nontrivial dispersive band,
although the mathematical theorem prohibits the fully gapped flat bands with finite Chern number in the finite-range hopping model. 

\subsection{Molecular-orbital Haldane model}
Another representative model of Chern insulators on a honeycomb lattice 
is the Haldane model~\cite{Haldane1988}. 
The MO-Haldane model reads
\begin{eqnarray}
\mathcal{H} = \sum_{\bm{k}} 
\left[ C^{\bigtriangleup, \dagger}_{\bm{k}}, C^{\bigtriangledown, \dagger}_{\bm{k}} \right]
h^{(\mathrm{H})}_{\bm{k}} 
\left[ 
\begin{array}{c}
C^{\bigtriangleup}_{\bm{k}} \\ 
C^{\bigtriangledown}_{\bm{k}}  \\
\end{array}
\right],
\end{eqnarray}
where
\begin{eqnarray} 
h^{(\mathrm{H})}_{\bm{k}}  = \epsilon_{\bm{k}}  \tau_0 + \bm{R}_{\bm{k}} \cdot \bm{\tau} + M \tau^z,
\end{eqnarray}
with $\tau_0$ being a $2\times2$ identity matrix, $\bm{\tau} = (\tau^x, \tau^y, \tau^z)$ being the Pauli matrices, 
\begin{eqnarray}
\epsilon_{\bm{k}} =2  t^{\prime} \cos \phi  \left[\cos \bm{k} \cdot \bm{a}_1 +  
\cos \bm{k} \cdot  \bm{a}_2 + \cos \bm{k} \cdot \left( \bm{a}_2 -\bm{a}_1 \right) \right], \nonumber \\
\end{eqnarray} 
\begin{eqnarray}
R^x_{\bm{k}} - iR^y_{\bm{k}} = t \sum_{p=1}^{3} e^{-i\bm{k} \cdot \bm{\delta}_p},  \label{eq:Rxy}
\end{eqnarray}
and  
\begin{eqnarray}
R^z_{\bm{k}} = 
2 t^{\prime}  \sin  \phi \left[\sin \bm{k} \cdot \bm{a}_1 - \sin \bm{k} \cdot  \bm{a}_2 + \sin \bm{k} \cdot \left( \bm{a}_2 -\bm{a}_1 \right) 
\right]. \nonumber \\
\label{eq:Rz}
\end{eqnarray}
In Eq. (\ref{eq:Rxy}), we have used
$\bm{\delta}_1 = \bm{a}_1 -\bm{a}_2$, $\bm{\delta}_2 =0$, and $\bm{\delta}_3 = \bm{a}_1$.
The analytical expression of the dispersion relations of two dispersive bands 
can be obtained by mapping the original eigenvalue problem of the $3 \times 3$ matrix to 
that of the $2 \times 2$ matrix, $h_{\bm{k}}^{(\mathrm{H})}\Psi_{\bm{k}}^\dagger \Psi_{\bm{k}}$~\cite{Hatsugai2011,Mizoguchi2019,Mizoguchi2018,Mizoguchi2019_2}.
The dispersion relations thus obtained are
\begin{widetext}
\begin{eqnarray}
\varepsilon_{\pm}(\bm{k}) = 
3\epsilon_{\bm{k}} + \frac{(R_{\bm{k}}^{x})^2 +(R^y_{\bm{k}})^2}{t} \pm
\sqrt{\left(9-\frac{(R_{\bm{k}}^{x})^2 +(R^y_{\bm{k}})^2}{t^2} \right) \left(R^z_{\bm{k}}+M \right)^2 +[(R_{\bm{k}}^{x})^2 +(R^y_{\bm{k}})^2]\left( 3 + \frac{\epsilon_{\bm{k}}}{t}\right)^2}.
\end{eqnarray}
\end{widetext} 

We plot the band structure and the Chern numbers 
for $(t,t^\prime,M,\phi) = (-1, -0.3, 0,\frac{\pi}{2})$ 
in Fig.~\ref{fig4}(a).
The zero-energy flat band is located between the upper and the lower dispersive bands. 
The lower dispersive band has the Chern number $-1$, 
thus the sum of the Chern numbers over the flat band and the upper dispersive band is $1$. 
Therefore, we can again realize the flat band touching the topologically nontrivial band.

We also compute the dispersions for the cylinder geometry, shown in 
Fig.~\ref{fig4}(b).
We see the chiral edge modes appear, 
due to the non-trivial Chern number and the bulk-edge correspondence~\cite{Hatsugai1993}.
Interestingly, the chiral edge modes crosses with the bulk flat band at zero energy. 
Note that similar dispersion is found in a topological flat-band model on a Lieb lattice~\cite{Weeks2010,Jana2019}.
\begin{figure}[t]
\centering
\includegraphics[clip,width=0.95\linewidth]{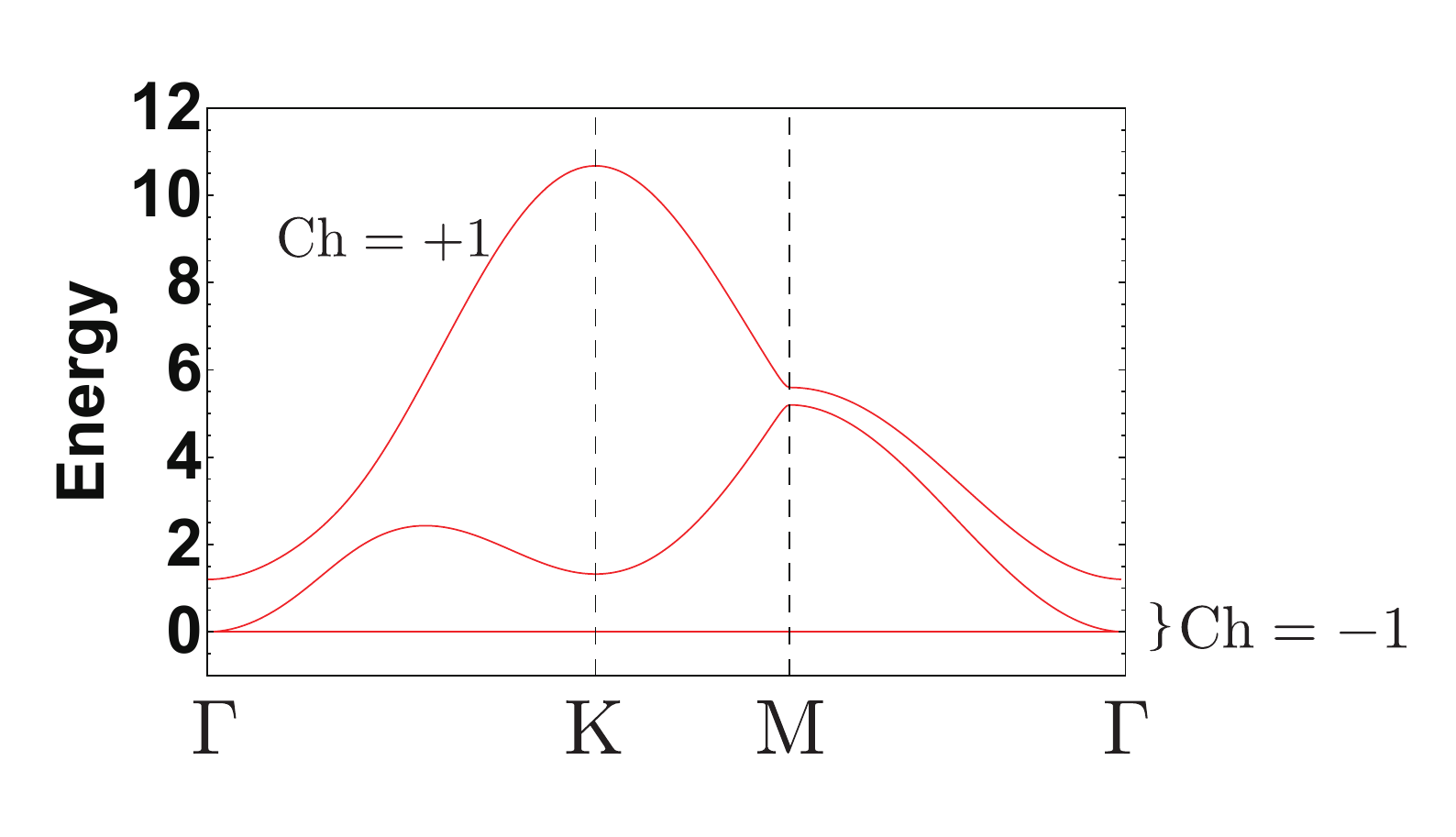}
\caption{ 
The band structure of the model of (\ref{eq:Ham_Haldane_prime})
with $(t,t^\prime,M,M^\prime,\phi) = (-0.6, -0.3, 0, 2, \frac{\pi}{2})$.
} 
\label{fig5}
\end{figure}

In Fig. ~\ref{fig4}(c), we map the Chern number of the lowest band in the $M$-$\phi$ space. 
We see that two topological phases with the Chern number 1 and $-1$ are separated by the gapless region,
where the lowest dispersive band overlaps with the flat band. 

In the MO Haldane model, the flat band is located between two dispersive bands.
For realization of topologically-nontrivial many-body states, on the other hand, 
it is often desirable to construct a model where the flat band has the lowest energy~\cite{Tang2011,Sun2011,Neupert2011,Sheng2011}.
In the present model, such a situation can be realized by considering a slightly-modified model:
\begin{eqnarray}
\mathcal{H} = \sum_{\bm{k}} 
\left[ C^{\bigtriangleup, \dagger}_{\bm{k}}, C^{\bigtriangledown, \dagger}_{\bm{k}} \right]
\tilde{h}^{(\mathrm{H})}_{\bm{k}} 
\left[ 
\begin{array}{c}
C^{\bigtriangleup}_{\bm{k}} \\ 
C^{\bigtriangledown}_{\bm{k}}  \\
\end{array}
\right],
\end{eqnarray}
with 
\begin{eqnarray} 
\tilde{h}^{(\mathrm{H})}_{\bm{k}}  =h^{(\mathrm{H})}_{\bm{k}} + M^\prime \tau_0. \label{eq:Ham_Haldane_prime}
\end{eqnarray}
The second term of (\ref{eq:Ham_Haldane_prime}) is an ``on-site" term for the MOs,
but it does not give rise to an entire shift of energy in the original kagome model~\cite{Hatsugai2011,Mizoguchi2019}.
In Fig.~\ref{fig5}, we show the band structure for the representative choice of parameters.
We see that the desirable band structure is obtained;
namely the flat band, touching the dispersive band, has the lowest energy, and the total Chern number 
for these two bands is indeed finite. 

\subsection{Molecular-orbital Kane-Mele model}
\begin{figure*}[t]
\centering
\includegraphics[clip,,width=0.95\linewidth]{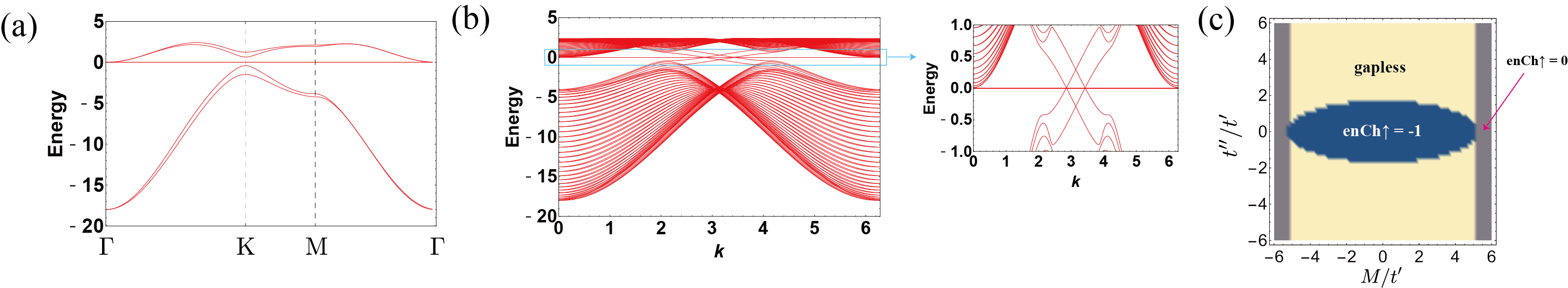}
\caption{ 
(a) The band structure of the MO Kane-Mele model for 
$(t,t^\prime,t^{\prime \prime},\phi,M) = (-1, -0.06, -0.05,\frac{\pi}{2},-0.1)$, 
(b) the same parameter on a cylinder geometry.
(c) The $t^{\prime \prime}$-$M$ phase diagram of the MO Kane-Mele model at $t^\prime = -0.06$.} 
\label{fig6}
\end{figure*}
Finally, we present an example of a model 
having $\mathbb{Z}_2$ topology.
To be concrete, we implement the Kane-Mele model~\cite{Kane2005} as the Hamiltonian of MOs. 
We now consider the spinful fermions.
The Hamiltonian is given as 
\begin{eqnarray}
\mathcal{H} = \sum_{\bm{k}} 
\left[ C^{\bigtriangleup, \dagger}_{\bm{k},\uparrow}, C^{\bigtriangledown, \dagger}_{\bm{k},\uparrow},
C^{\bigtriangleup, \dagger}_{\bm{k},\downarrow}, C^{\bigtriangledown, \dagger}_{\bm{k},\downarrow} \right]
h^{(\mathrm{KM})}_{\bm{k}} 
\left[ 
\begin{array}{c}
C^{\bigtriangleup}_{\bm{k},\uparrow} \\ 
C^{\bigtriangledown}_{\bm{k},\uparrow}  \\
C^{\bigtriangleup}_{\bm{k},\downarrow} \\ 
C^{\bigtriangledown}_{\bm{k},\downarrow}  \\
\end{array}
\right], \nonumber \\
\end{eqnarray}
where
\begin{widetext}
\begin{eqnarray}
h^{(\mathrm{KM})}_{\bm{k}}  = 
\left(
\begin{array}{cccc}
\epsilon_{\bm{k}}(\phi) + R^z _{\bm{k}}(\phi) + M & R^{x}_{\bm{k}}  -iR^{y}_{\bm{k}}  & 0 & t^{\prime \prime} \alpha_{\bm{k}} \\
R^{x}_{\bm{k}}  + iR^{y}_{\bm{k}}& \epsilon_{\bm{k}}(\phi) -R^z_{\bm{k}}(\phi)  -M & t^{\prime \prime} \beta_{\bm{k}} & 0\\
0 &  t^{\prime \prime} \beta_{\bm{k}}^\ast& \epsilon_{\bm{k}}(-\phi) + R^z _{\bm{k}}(-\phi)+M &  R^{x}_{\bm{k}} -iR^{y}_{\bm{k}}  \\
t^{\prime \prime} \alpha_{\bm{k}}^\ast & 0 & R^{x}_{\bm{k}} + iR^{y}_{\bm{k}}&  \epsilon_{\bm{k}}(-\phi) - R^z _{\bm{k}}(-\phi)-M \\ 
\end{array}
\right),
\end{eqnarray}
\end{widetext}
with $t^{\prime \prime}$ being
the Rashba spin-orbit coupling; the explicit forms of $\alpha_{\bm{k}}$ and $\beta_{\bm{k}}$ are
\begin{eqnarray}
\alpha_{\bm{k}}= i  \left[ e^{i\frac{2\pi}{3}} + e^{-i \bm{k} \cdot (\bm{a}_1 -\bm{a}_2) } 
+ e^{-i \bm{k} \cdot \bm{a}_1  + \frac{i 4\pi}{3}} \right], \nonumber \\
\end{eqnarray}
and 
\begin{eqnarray}
\beta_{\bm{k}}= i  \left[ e^{-i\frac{ 2\pi}{3}} + e^{-i \bm{k} \cdot (\bm{a}_1 -\bm{a}_2) } 
+ e^{-i \bm{k} \cdot \bm{a}_1 - \frac{i 4\pi}{3}} \right], \nonumber \\
\end{eqnarray}
In the following, we set $\phi = \frac{\pi}{2}$, for simiplicity.

We plot the band structure for $(t,t^\prime,t^{\prime \prime},M) 
= (-1, -0.3, -0.5, 0)$ in Fig.~\ref{fig6}(a).
The doubly-degenerate flat bands touch the dispersive bands at the $\Gamma$ point.
Figure~\ref{fig6}(b) shows the dispersion relation on a cylinder geometry. 
We see that the helical edge states are intersected by the bulk flat bands. 

In Fig.~\ref{fig6}(c), we depict the phase diagram of this model,
obtained by calculating the entanglement Chern number~\cite{Fukui2014,Araki2016,Fukui2016,Araki2017},
enCh$^\sigma$, for the lowest two bands.
To be concrete, the entanglement Chern number is defined for 
the eigenstates of 
the entanglement Hamiltonian 
$H_{\rm en}(\bm{k})$: 
\begin{eqnarray}
H_{\rm en}(\bm{k})^{\rm T} = 
\ln \left[\frac{1-P_{\uparrow} P_{-}(\bm{k}) P_{\uparrow}}{P_{\uparrow} P_{-}(\bm{k}) P_{\uparrow}}\right],
\end{eqnarray} 
with 
\begin{eqnarray}
P_{\uparrow} =\mathrm{diag}
\left(1,1,1, 0,0,0\right),
\end{eqnarray}
and 
\begin{eqnarray}
P_{-}(\bm{k}) = \sum_{n=1}^2 \bm{u}_{\bm{k},n} \bm{u}^{\dagger}_{\bm{k},n}.
\end{eqnarray}
Here $\bm{u}_{\bm{k},n}$ denotes the $n$-th eigenvector of $\mathcal{H}_{\bm{k}} = \Psi_{\bm{k}} h^{(\mathrm{KM})}_{\bm{k}} \Psi^\dagger_{\bm{k}}$.
In Fig.~\ref{fig6}(c), we find three phases: 
the $\mathbb{Z}_2$ topological phase (enCh$^\sigma = 1$),
the trivial phase (enCh$^\sigma = 0$),
and the gapless phase where one of the lower dispersive bands overlaps with the flat bands.

\section{Summary and outlook \label{sec:summary}}
In summary, we have introduced
a systematic method to construct topological flat-band models 
with finite-range hoppings. 
The existence of flat bands is guaranteed since the model is constructed by the MOs.
Although flat bands themselves are not allowed 
to have non-trivial topological numbers, the dispersive bands 
touching the flat bands 
can have the topologically nontrivial nature.
Such models will serve as a platform to look for intriguing phenomena arising from the topology and the correlation effects.
In this light, studying the many-body effects in these models will be an interesting future problem. 

Throughout this paper, we consider the models on a two-dimensional kagome lattice, 
but it is straightforward to apply our method to other lattices, including ones in three dimensions. 
For instance, if we consider the model on a pyrochlore lattice, 
the MOs are defined on a diamond lattice.
Thus, to find the $\mathbb{Z}_2$ topological model, the Fu-Kane-Mele model~\cite{Fu2007} can be used 
as a MO Hamiltonian; in the Appendix~\ref{sec:pyrochlore}, we explicitly construct such a model.
If we consider the many-body effects in the model thus obtained, 
the flat band will lead to the ferromagnetism, while the $\mathbb{Z}_2$ topological nature leads to 
the topological surface states.
Then, the interplay between these two may lead to the quantum anomalous Hall effect, 
as in the case of magnetic topological insulators~\cite{Chang2016,Tokura2019}. 

\acknowledgements
We thank T. Yoshida for fruitful discussions. 
This work is supported by JSPS KAKENHI, Grants No.
JP17H06138 and No. JP16K13845 (YH), MEXT, Japan.

\appendix
\section{A model in three dimensions: MO Fu-Kane-Mele model \label{sec:pyrochlore}}
\begin{figure}[tb]
\centering
\includegraphics[clip,,width=0.93\linewidth]{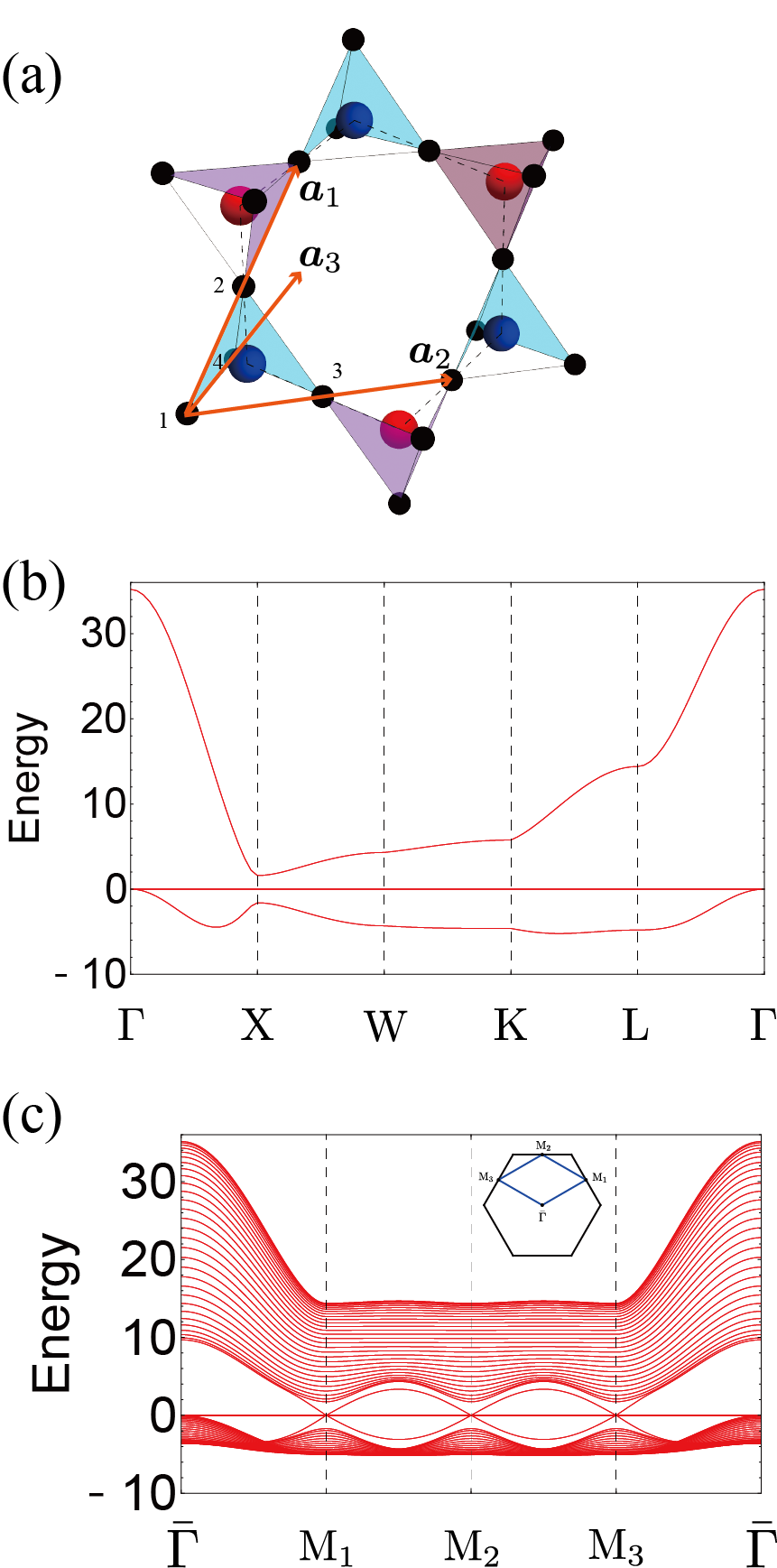}
\caption{ 
(a) The pyrochlore lattice. Orange arrows represent the lattice vectors:
$\bm{a}_1 = (0,1/2,1/2)$, $\bm{a}_2 = (1/2,0,1/2)$, and $\bm{a}_3 = (1/2,1/2,0)$.
Blue and red spheres represent the sites of the diamond lattice 
on which the MOs for the upward and the downward tetrahedra are defined, respectively.
(b) The band structure of the MO Fu-Kane-Mele model for $(t,t^\prime,\delta t) = (1,0.125,0.4)$. 
The high-symmetry points in the Brillouin zone are $\Gamma=(0,0,0)$,
$X= (2 \pi,0,0)$, 
$W= \left(2 \pi, \pi, 0\right)$,
$K=\left(\frac{3\pi}{2},\frac{3\pi}{2},0 \right)$,
and L$=\left( \pi, \pi, \pi \right)$. 
(c) The band structure for the same parameters with a slab geometry. 
The surface is perpendicular to $[1\bar{1} \bar{1}]$ direction. 
The high-symmetry points in the surface Brillouin zone are
$\bar{\Gamma}=(0,0)$, $M_{1} = \left( \sqrt{2} \pi, \frac{\sqrt{2} \pi}{\sqrt{3}} \right)$, 
$M_{2} = \left( 0, \frac{2 \sqrt{2} \pi}{\sqrt{3}} \right)$, and 
$M_{3} = \left( - \sqrt{2} \pi, \frac{\sqrt{2} \pi}{\sqrt{3}} \right)$. 
 } 
\label{fig7}
\end{figure}
In this Appendix, we present an example of the $\mathbb{Z}_2$ 
topological insulator with exact flat bands on a pyrochlore lattice.
The pyrochlore lattice has a corner-sharing network of tetrahedra. 
Defining the MO on each tetrahedron, one can find that it forms the diamond lattice.
To be concrete, the MO on an upward tetrahedron [colored in cyan in Fig.~\ref{fig7}(a)] at the unit cell $\bm{R}$ is
\begin{subequations}
\begin{eqnarray}
C_{\bm{R},\rm{U},\sigma} = c_{\bm{R},1,\sigma} + c_{\bm{R},2,\sigma} +c_{\bm{R},3,\sigma}+c_{\bm{R},4,\sigma},
\end{eqnarray}
and that on a downward tetrahedron [colored in magenta in Fig.~\ref{fig7}(a)] is
\begin{eqnarray}
C_{\bm{R},\rm{D},\sigma} = c_{\bm{R},1,\sigma} + c_{\bm{R}-\bm{a}_1,2,\sigma} +c_{\bm{R}-\bm{a}_2,3,\sigma}+c_{\bm{R}-\bm{a}_3,4,\sigma}, \nonumber \\
\end{eqnarray}
where $c_{\bm{R},a,\sigma}$ represents the annihilation operator defined on a site of the pyrochlore lattice
with at the unit cell $\bm{R}$ and the sublattice $a$ having spin $\sigma$.

Following the method discussed in the main text, 
we implement a model for three-dimensional topological insulators
as a Hamiltonian for the MOs.
Specifically, we employ the Fu-Kane-Mele model~\cite{Fu2007}:
\end{subequations}
\begin{eqnarray}
\mathcal{H} &=& \sum_{\langle I,J \rangle, \sigma} t_{I,J} C_{I,\sigma}^\dagger C_{J,\sigma} + (\mathrm{H.c.}) \nonumber \\
&+& 8 i t^\prime \sum_{\langle \langle  I,J \rangle \rangle,\sigma,\sigma^\prime} [ \bm{\tau} \cdot \bm{d}^{(1)}_{I,J} \times \bm{d}^{(2)}_{I,J}  ]_{\sigma,\sigma^\prime} C^\dagger_{I,\sigma}C_{J,\sigma^\prime}, \nonumber \\
\end{eqnarray}
where $I$ and $J$ represent the sites on a diamond lattice on which the MOs are placed, 
and $\langle, \rangle$ and $\langle \langle, \rangle \rangle$
represent, respectively, the nearest neighbor and the next nearest neighbor pairs.
$\bm{d}^{(1)}_{I,J}$ and $\bm{d}^{(2)}_{I,J}$ are the nearest neighbor bonds which traverse the sites $I$ and $J$.  
For the nearest-neighbor hopping $t_{I,J}$, we set $t_{I,J} = t + \delta t$ if $\bm{r}_J -\bm{r}_I$ is parallel to $\bm{a}_1$, and $t_{I,J} = t $ otherwise; $\bm{r}_I$ and $\bm{r}_J$ represent the position of $I$ and $J$, respectively.

In Fig.~\ref{fig7}(b), we plot the band structure for the representative set of parameters.
The band structure has many things in common with the kagome models discussed in the main text;
there is a zero-energy flat band, which has four-fold degeneracy, 
and two dispersive bands, each of which has two-fold degeneracy.
Further, the lower dispersive band touches the flat band at the $\Gamma$ point.

To verify the topological nature, we plot the band structure for the slab geometry in Fig.~\ref{fig7}(c).
We clearly see the topologically protected surface states, penetrating the bulk flat band. 
This is another common feature to the kagome-lattice model.
As we have mentioned, such a band structure may be a platform of 
the quantum anomalous Hall effect induced by the flat-band ferromagnetism, 
when the interaction is turned on.

\end{document}